\documentclass[12pt, a4paper, oneside, openany]{article}
\usepackage[a4paper, lmargin=0.1\paperwidth, rmargin=0.17\paperwidth, tmargin=0.1111\paperheight, bmargin=0.1111\paperheight]{geometry} 

\usepackage[english]{babel}
\usepackage{parskip}
\usepackage{rotating}
\usepackage{amssymb,mathtools}
\usepackage{helvet} 
\usepackage{graphicx}
\usepackage{tabularx, booktabs, multirow}
\usepackage{ragged2e}
\newcolumntype{R}{>{\RaggedRight\arraybackslash}X}
\usepackage[justification=raggedright]{caption}
\usepackage[stable,hang,flushmargin]{footmisc} 
\usepackage{xcolor}
\usepackage{csquotes} 
\usepackage[pages=some]{background}
\usepackage{fancyhdr}
\usepackage[sectionbib]{natbib}
\usepackage{chapterbib} 

\usepackage{longtable}
\usepackage{graphicx,dcolumn,floatpag}
\usepackage{algorithm}
\usepackage{algorithmic}
\usepackage[inline]{enumitem}
\usepackage{svg}
\usepackage{float}
\usepackage{array}
\usepackage{xurl}
\usepackage{mathptmx}%
\usepackage{ltablex}
\usepackage{makecell}
\usepackage[caption=false]{subfig}
\usepackage{tablefootnote}
\usepackage{imakeidx}

\usepackage{acro}
\DeclareAcronym{DLT}{
  short = DLT,
  long = distributed ledger technology,
  long-plural-form = distributed ledger technologies
}

\DeclareAcronym{TEA}{
  short = TEA,
  long =  triple-entry accounting
}  

\DeclareAcronym{TEB}{
  short = TEB,
  long =  triple-entry book-keeping
} 

\DeclareAcronym{STRs}{
  short = STRs,
  long =  Shared Transaction Repositories 
}  
 
\DeclareAcronym{DBT}{
  short = DBT,
  long =  Distributed Book Technology
}  
 
\DeclareAcronym{DJT}{
  short = DJT,
  long =  Distributed Journal Technology
} 
  
\DeclareAcronym{P2P}{
  short = P2P,
  long =  peer-to-peer
}
\DeclareAcronym{GLT}{
  short = GLT,
  long =  General Ledger for Transactions
}

\DeclareAcronym{GLR}{
  short = GLR,
  long =  General Ledger for Reporting
}

\DeclareAcronym{IDEA}{
  short = IDEA,
  long =  Immutable Double-entry Accounting
}

\DeclareAcronym{PTA}{
  short = PTA,
  long =  Plain Text Accounting
}

\PassOptionsToPackage{hyphens}{url}
\usepackage{hyperref} 
\hypersetup{colorlinks=true,
	linkcolor=blue,
	citecolor=blue,
	urlcolor=blue}

\addto\extrasenglish{%
}
\addto\extrasenglish{%
}	

\addto\extrasenglish{%
}

\usepackage[clearempty]{titlesec}

\newcommand{\chapterbib}{%
\bibliographystyle{apalike}
\bibliography{references}%
}

\backgroundsetup{
scale=1,
color=black,
opacity=1,
angle=0,
contents={%
  \includegraphics[height=\paperheight]{cover.jpg}
  }%
}
\title{Digital Assets: Pricing, Allocation and Regulation}
\date{}

\titleformat{\subsection}
{\normalfont\fontfamily{phv}\fontsize{12}{17}\bfseries\itshape}{\thesubsection}{1em}{}

\begin{document}
\title{Triple-entry Accounting, Blockchain and Next of Kin: Towards a Standardisation of Ledger Terminology}
\date{}
\maketitle
\vspace{-1cm}

\begin{center}
\author{%
  Juan Ignacio Ibañez\footnotemark[1]%
  \and\hspace{0.25cm}
  Chris N. Bayer\footnotemark[2]%
  \and\hspace{0.25cm}
  Paolo Tasca\footnotemark[1]%
  \and\hspace{0.25cm}
  Jiahua Xu\footnotemark[1]%
}
\footnotetext[1]{Centre for Blockchain Technologies, University College of London and Exponential Science}
\footnotetext[2]{Development International e.V.}
\end{center}

\vspace{1.5cm}

\begin{abstract}
\Acf{TEA} is simultaneously a novel application in the blockchain universe and one of the many concepts applied in blockchain technology. Its Wild Wild West status is accompanied by a lack of consistent and comprehensive set of categories, a state of play that impedes a proper apprehension of the technology, leading to contradictions and oversight of important nuances. To clearly delineate the confines of \ac{TEA} within the world of blockchain, we provide building blocks to standardise its terminology. Particularly, we distinguish between essential elements such as accounting and bookkeeping\index{bookkeeping}, as well as between decentralised systems, distributed ledgers and distributed journals.
\end{abstract}



\section{Introduction}
\ac{TEA} is one of the most innovative concepts of the past decades. It introduces a shared transaction record that constitutes a single source of truth, which is secured through triple-signed receipts and is usually maintained by means of blockchain. This idea was first introduced through Ian Grigg’s Ricardo Payment System beginning in 1995. In 1998, Todd Boyle independently started to develop \ac{STRs}, with some influence from William E. McCarthy’s REA accounting model. In 2004, Grigg and Boyle converged, and the latter’s work was integrated onto the former’s \citep{Ibanez2023REASystems}.

While \ac{TEA} is extremely innovative, \enquote{there is a dearth in academic research on the topic (\ac{TEA}), which is extremely limited} \citep{cai2019triple}. This includes attempts to specify the meaning of the terms at use. Meanwhile, practitioners appear not to be putting much effort into a comprehensive conceptualisation of \ac{TEA} and related terms. In turn, they focus on the technical aspects of their product, while taking the vocabulary for granted. This is illustrated by the fact that most companies developing or commercialising \ac{TEA} projects rarely specify the terms used in their whitepapers.

The concept of \ac{TEA} is thus rarely defined. It appears, however, that it is often assumed to mean a shared ledger – and, if blockchain technology is used, it is advertised as a decentralised shared ledger. Nonetheless, were \ac{TEA} equivalent to a shared ledger, there would be no need for the concept of \ac{TEA} in the first place, which suggests instead that a \ac{TEA} system is a sub-type of sacred ledger.

Matters are complicated further by the question of whether Bitcoin\index{Bitcoin} is a \ac{TEA} system. It is often described as one \citep{Grigg2011, tyra2014triple} but the reason therefor is unclear. Furthermore, were this to be the case, it would pose the question of whether all other cryptocurrencies\index{cryptocurrency} are also \ac{TEA} use cases. If they were all also \ac{TEA} use cases, one would be left to wonder what distinguishes those products specifically advertised as \ac{TEA}.

Consider furthermore that it is speculated that \ac{TEA} was one of the influences behind Bitcoin\index{Bitcoin} itself \citep{Ibanez2023REASystems}, that is, behind the use case introducing blockchain technology as such. This raises the question of whether there is anything intrinsically \ac{TEA}-like in blockchain technology, meaning that blockchains would constitute \ac{TEA} by definition and that the term \enquote{blockchain-based \ac{TEA} systems} would be redundant, or the two concepts are not in perfect overlap.

This leads to the enquiry of what blockchain is, as well as what (if any) difference lies with the concept of distributed ledger technology or \ac{DLT}. Nevertheless, these notions are also elusive: the prime characteristic of a \ac{DLT} is not that it is distributed, as distributed systems need not be \ac{DLT}. Many maintain that what distinguishes distributed ledger technology from other shared ledger systems is not that \ac{DLT} is distributed, but that it is decentralised \citep{fresno2018cual,TradeIX2018}. However, a number of \acp{DLT} are not fully decentralised in every possible way. Bitcoin\index{Bitcoin}, for instance, displays centralisation in its mining pools \citep{Kimani2018BlockchainGrigg}. Nevertheless, it would be hard to argue that this makes Bitcoin\index{Bitcoin} any less of a \acp{DLT}.

A number of papers have attempted to build a taxonomy for blockchain/\ac{DLT} systems \citep{ballandies2018decrypting, beinke2018towards, garay2020sok, glaser2015beyond, Ibanez2023REASystems, ismail2019towards, labazova2019hype, mohsin2019blockchain, sarkintudu2018taxonomy, tasca2019taxonomy, walsh2016new}. Nevertheless, these works tend to focus on organising existing and possible systems into typologies, usually taking the higher-level categories for granted\footnote{For an exception, see \cite{rauchs2018distributed}}.

For the reasons explained above, there is a deficit in this field of research. Existing categories are insufficient to give a proper account of \ac{TEA}. This includes extant blockchain taxonomies.

This chapter sets out to fill in this gap. We build on existing research that describes the taxonomy and landscape of accounting and \ac{DLT}. We find that an adequate terminology to give account of \ac{TEA} requires the introduction of accounting vocabulary, not just for \ac{TEA}, but also for \ac{DLT} in general. 

With this goal in mind, this chapter is structured in the following manner: first, we lay out the methods employed. Second, we present a background discussion of the distinction between distributed systems and decentralised systems, as well as the one between accounting and book-keeping\index{bookkeeping}, and journals and ledgers. Third, we define \ac{TEA} and specify the architecture of a \ac{TEA} system. Fourth, based on the previous steps, we attempt to standardise terminology, to clarify misunderstandings and set the stage for further analysis. Fifth, we conclude the chapter with summary observations and recommendations for further research.

\section{Methodology}
Approaching the topic from the perspective of accounting theory and practice, this chapter reviews usage in \ac{TEA} use cases identified in \cite{ibanez2021efficiency} by consulting their whitepapers. It further elaborates on the categories relevant to \ac{TEA} and of the architectural requirements and possibilities of a \ac{TEA} system. This is done, first, through a literature review. The literature consulted, quoted throughout the chapter, notably covers the disciplines of accounting and computing in general, and of \ac{TEA}, cryptography, \ac{DLT} and blockchain in particular. Second, we conduct key informant interviews per phone or email correspondence with salient individuals in the world of \ac{TEA}. The interviewees are listed in \autoref{acknow}.

\section{Results}
Our research yielded the following findings: the coherent usage of accounting vocabulary is necessary for a proper apprehension of \ac{TEA}, \ac{DLT} systems and blockchain. It has been often stated that \enquote{blockchain is fundamentally an accounting technology} \citep{ICAEW2018}, yet not only have the main usages of the blockchain so far, for example, cryptocurrencies\index{cryptocurrency}– not been about accounting, but also accounting terminology has been used with excessive latitude until the present.

In this direction, we find that the specificity of \ac{DLT}-based \ac{TEA} systems can only be apprehended by applying the distinction, used in the accounting discipline, between book-keeping\index{bookkeeping} and accounting \citep{Ibanez2022Triple-entryBlockchain}. Furthermore, we find that the distinction between journals and ledgers as the two possible types of books is also important to sustain this specificity. Moreover, we find that a correct conceptualisation of the term \ac{DLT} requires preliminary awareness of series of contrasts.

This section contains a detailed presentation of our findings.

\subsection{Conceptualisation of \ac{TEA} in Whitepapers}
As explained in the Introduction, \ac{TEA} is a concept used in whitepapers, but seldom defined. This emerges from the consultation of the whitepapers of the \ac{TEA} use cases identified in \cite{ibanez2021efficiency}. For instance, The Accounting Blockchain’s whitepaper simply defines \ac{TEA} as the act of recording a transaction in the sets of books of the two parties to the transactions simultaneously and does no further attempt at conceptualisation \citep{AB2018}.

Similarly, Provenance Blockchain does not mention the concept of \ac{TEA} in its whitepaper (Figure Technologies), despite having been advertised as a \ac{TEA} use paper by board member \cite{Pompliano2019}. The same is the case of another \ac{TEA} use case: \cite{Auditchain2018}. In interviews conducted with Jason Meyers (founder of Auditchain) and Anthony Pompliano, both confirmed that the product put forward by their companies constituted \ac{TEA} use cases.

Similarly, Open Transactions’ whitepaper only states that \enquote{we proposed the use of triple-signed receipts} \citep{Odom2015}, without further explanation. Request’s whitepaper solely claims that \enquote{it represents a switch from double-entry accounting to triple-entry accounting} \citep{Request2018}, also without further explanation. This is also the case for Ledgerium’s whitepaper, which only mentions that \enquote{we utilise a decentralised ledger through a triple entry accounting system} \citep{Mehmood2019}.

In a similar fashion, Corda’s sole public explanation of the term is that the design was \enquote{inspired by previous work, including that (…) on triple entry accounting} (Brown: 17). Similarly, PayPie (2018: 5) only explains that \enquote{Adding blockchain record in the world of debit and credit effectively creates a triple entry accounting system.} Only \cite{Pacio2020} whitepaper contains some conceptual effort, with a detailed Standardised Semantic Information Model describing how its \ac{TEA} system would work, as well as prerequisites for a functional \ac{TEA} system

\subsection{Preliminary Concepts: Decentralisation, Distribution, Accounting and \ac{DLT}}
In this section, we point out the distinction between distributed and decentralised systems, together with the divide between (triple-entry) accounting and (triple-entry) book-keeping\index{bookkeeping}, and between \ac{DBT}, \ac{DJT} and \ac{DLT}. 

\subsubsection{The Distinction Between the Centralised, the Decentralised, and the Distributed}
\begin{figure}[H]
  \centering
  \includegraphics[width=\textwidth]{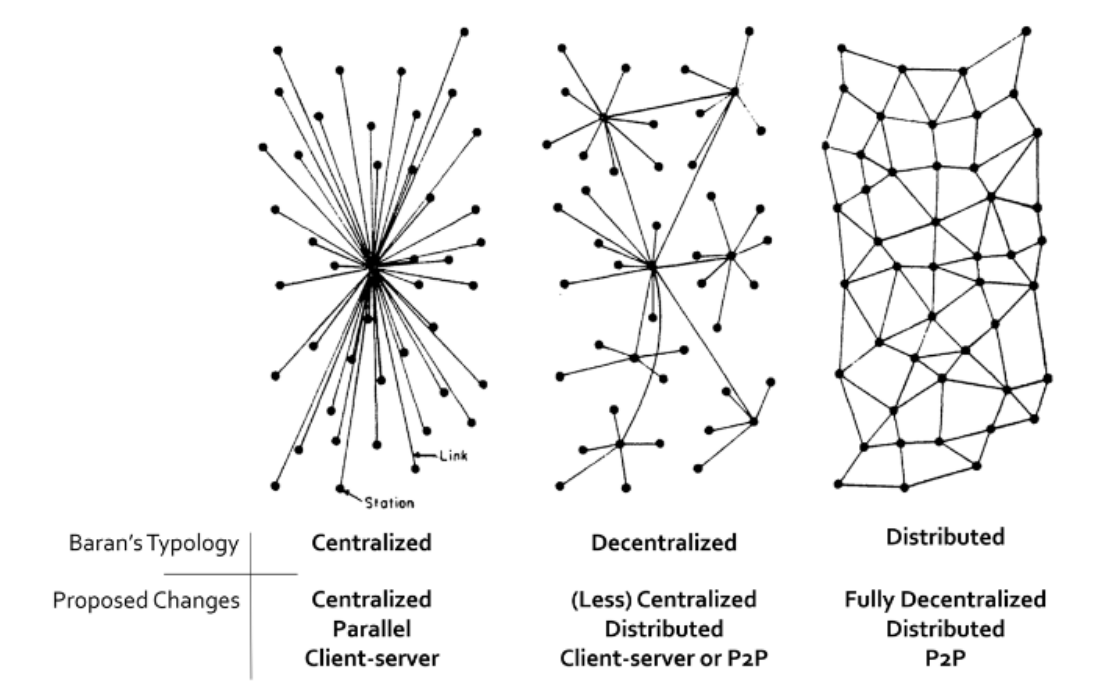}
  \caption{Centralised, decentralised and distributed networks according to \cite{baran1964distributed} and revised categories}
  \label{figureTEA.1}
\end{figure}

In 1964, \cite{baran1964distributed} claimed that there were two types of networks: \enquote{centralised (or star) and distributed (or grid or mesh)}. Baran also argued that \enquote{in practice, a mixture of star and mesh is used to form communication networks (...) a \enquote{decentralised} network.} Baran’s typology is shown in \autoref{figureTEA.1}.

This distinction has found wide adoption within the blockchain community, but results are anachronistic. Fundamentally, it does not match current use in computing, and thus creates a series of false impressions (see also \cite{Buterin2017}). In fact, the names applicable to the second and third diagrams should arguably be reversed \citep{grange2016mesh}.

Despite not capturing the distinction between systems that are centralised by design and systems that moved to a degree of centralisation from an originally decentralised design, Baran’s illustration of a centralised network is not controversial. However, the depiction of a decentralised network could be challenged, on the grounds that there is a centre to the network. Moreover, Baran appears to depict decentralisation as equivalent to deconcentration of functions. Although the fact that some nodes are shown as more connected than others is not problematic in itself, the depiction of decentralisation as a central node delegating connections to intermediate centres may lead to confusion.

Currently, a better understanding for the term \enquote{decentralised} is, as the name suggests, a system lacking a (decision-making and single point of failure)\footnote{According to Ethereum\index{Ethereum} co-founder Vitalik Buterin, to the two aforementioned dimensions of decentralisation (political and architectural, respectively) one should add a third one: logical decentralisation, namely whether the system’s interface looks monolithic or not. \enquote{One simple heuristic is: if you cut the system in half, including both providers and users, will both halves continue to fully operate as independent units?} \citep{Buterin2017}} centre. The Internet is an example of a decentralised system. While there is still no ultimate centre, the Internet is also not fully decentralised, as there are a number of decision-making centres or \enquote{hubs} within its structure. Cash transactions, by definition, are almost completely decentralised. Therefore, there is a gradient between the centralised and the fully decentralised, in the middle of which the partly decentralised exists.

Whether a system is distributed, in turn, is an entirely different affair. In a distributed computing system, each independent computer or node (with a local memory) communicates with the others through message-passing. This comes in opposition to the alternative of having one shared memory, that is, of being a \enquote{parallel system,} such that the collection of independent nodes appears to the users as a single computer \citep{van2016brief}. For instance, if the customer of a bank notices no difference between doing a transaction in one office of a bank and another one on the other side of the country (despite there being numerous computers in the system, plus a master computer per office and a central computer at the headquarters), it is a distributed system \citep{van2016brief}.

What Baran’s third diagram shows is not what is currently known as a distributed network, but a purely \ac{P2P} network. A distributed system may be centralised or decentralised. A \ac{P2P} system is a particular type of distributed system in which the clients also act as servers: they share their hardware resources for other users to access without the need for intermediaries, that is, they are \enquote{servants} (clients + servers) – in contrast to the client/server model. As a result, it is usually decentralised.

A \ac{P2P} system may have a central entity, but if this is the case it is known as hybrid \ac{P2P} system. Pure \ac{P2P} systems are decentralised: the nodes are equipotent and, thus, there are no hubs \citep{schollmeier2001definition}. File-sharing mechanisms like Napster and BitTorrent constitute examples of \ac{P2P} systems. Bitcoin\index{Bitcoin} is an example of a \ac{P2P} system that has ended up forming hubs (mining pools) and thus does not exhibit full decentralisation.

\begin{figure}[H]
  \centering
  \includegraphics[width=\textwidth]{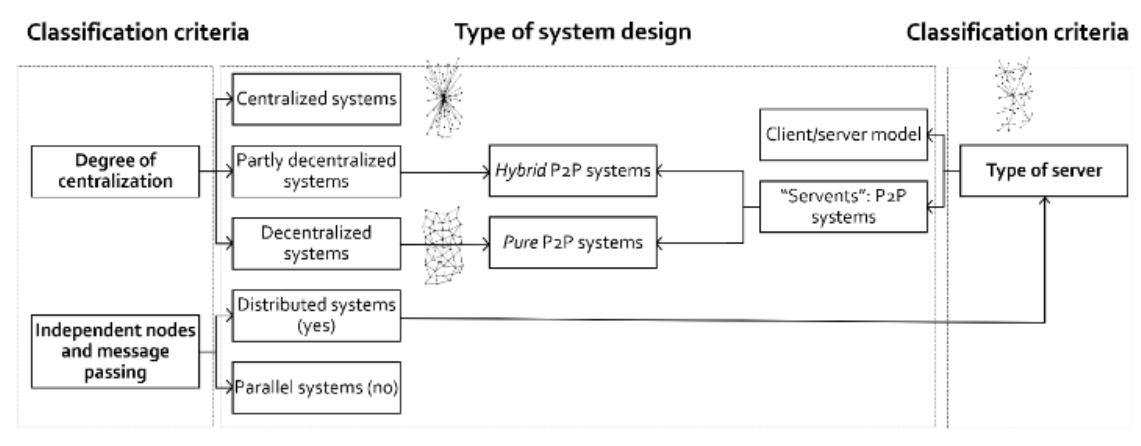}
  \caption{Whether a computing system is centralised, decentralised, distributed, or \acf{P2P} constitutes in each case an answer to a different question. The categories should thus not be construed as a continuum}
  \label{figureTEA.2}
\end{figure}

In conclusion, as shown in \autoref{figureTEA.2}, whether a system is decentralised (or centralised), distributed (or parallel), and/or \ac{P2P} (or client/server) depends on the answer to different questions. The anachronism given by the application of Baran’s typology misconstrues this as if \enquote{centralised,} \enquote{decentralised} and \enquote{distributed} were three possible answers to one common question, which is not the case.

\subsubsection{The Difference Between Book-keeping and Accounting}
This chapter distinguishes types of triple-entry systems based on whether they merely keep a shared record of transactions, or whether they undertake a more complex processing task on the basis of that information. For this purpose, the distinction between book-keeping\index{bookkeeping} and accounting is used \citep{Ibanez2023REASystems, Ibanez2022Triple-entryBlockchain}. However, we should note that the meanings assigned to recordkeeping, book-keeping\index{bookkeeping}, and accounting in the literature are not completely homogeneous. 

This dispute is difficult to settle because the peer-reviewed literature has not directly dealt with this question. At best, published papers spend no more htan one sentence on each term, whereas handbooks and manuals provide definitions outright without justifying them.

While some believe that there is no difference between book-keeping\index{bookkeeping} and accounting \citep{Ibanez2023REASystems}, most do. \cite{edwards1960early}, for example, claims that recordkeeping would be to preserve the documentary evidence of a particular transaction. Book-keeping\index{bookkeeping}, in turn, would be a historical stage that came after recordkeeping and consists of \enquote{analysing, classifying, and recording transactions.} Finally, the ulterior stage of accounting is \enquote{bookkeeping\index{bookkeeping} with additional refinements of financial summarisation and a control function added.}

However, the majority position seems to be that book-keeping\index{bookkeeping} and (\enquote{mere}) recordkeeping are synonyms, whereas accounting \enquote{builds on top of bookkeeping\index{bookkeeping} to make that information flow to the decision-making areas of the firm, by means of systematising, compiling, collating, synthesising, processing, analysing and/or auditing} \citep{Ibanez2023REASystems, Ibanez2022Triple-entryBlockchain}. This position is found in the peer-reviewed literature \citep{rukhiran2018house, vollmer2003bookkeeping} and in many instruction manuals and handbooks \citep{jnr1977visible, Ge2005, mutero2017introduction, peters2011introduction, Wild2011}. This is the distinction that we follow\footnote{In practice, this differentiation still leaves some room for debate. One may argue over where the line between accounting and bookkeeping\index{bookkeeping} lies, since e.g. when recording a transaction, some classification may be involved, but only through the mechanical application of rules discovered or invented in the accounting profession.}. 

We find that the suggestion by \cite{Grigg2017b, Grigg2019} - see also \cite{Ibanez2023REASystems, Ibanez2022Triple-entryBlockchain} - of importing this distinction to the classification of triple-entry systems allows to apprehend their differences better. It is thus recommendable to differentiate between \ac{TEB} and \ac{TEA}. \ac{TEB} systems simply record transactions in a triple-entry fashion (e.g. Bitcoin\index{Bitcoin}, see \cite{Grigg2011, Ibanez2023REASystems}) and \ac{TEA} systems involve an accounting software built on a layer of \ac{TEB} \citep{Ibanez2022Triple-entryBlockchain}.

\subsubsection{An Accounting Typology for Blockchain-inspired Technology}
With the aforementioned categories clarified, we can proceed to establish a typology that functions as a proper \enquote{toolbox} to give an account of the \ac{TEA} phenomenon. Before explaining triple-entry recordkeeping, we need to understand recordkeeping itself.

Recordkeeping is also known as book-keeping\index{bookkeeping} because the instrument where a firm’s transactions with other companies are recorded is called a \enquote{book.} The book where transaction raw data is first recorded – sequentially – is called a journal. Accounting records are different from book-keeping\index{bookkeeping} records in that transactions are recorded analytically, rather than just sequentially: transactions are classified so that the resulting record shows information that is meaningful for business life (for instance, it facilitates decision-making or financial reporting). Thus, the accounting process takes the information of the journal and posts it in a second book, where information is organised analytically. This book is known as ledger. In consequence, \textit{accounting happens in ledgers, whereas book-keeping\index{bookkeeping} is limited to journals.}

If the two companies engaging in a transaction use the same book, it is known as a shared book. It is hard to conceive a shared book on paper. Rather, shared books are generally held in computer systems common to (or connected to) both firms. Computer systems may be parallel or distributed. If the book is held on a distributed system, it is a \textit{distributed book}. If, as a consequence of being unable to trust each other (i.e., being \textit{adversarial}), the computers (nodes) in the system resort to verification techniques to maintain a consensus\index{consensus} \footnote{Trust and consensus\index{consensus} solve are ways to implement the \enquote{what you see is what I see} principle, known by the WYSIWIS acronym.} about the transaction history, the technology for this distributed book receives a particular industry name: \ac{DBT}\footnote{For examples of usage of the term \ac{DBT}, see \cite{Liu2018AnBlockchain}.}. 

If the \ac{DBT} record is a journal, the system may be called \ac{DJT}. If it is ledger, \ac{DLT}.\footnote{For works advocating a proper usage of the terms \ac{DLT} and \ac{DJT}, see \cite{rauchs2018distributed}, \cite{UniversityofDerby2019DistributedTechnologies} and \cite{Shea2015TheJournal}.} Hence, under a proper usage of accounting terminology, and because because \textit{not all books are ledgers}, \ac{DLT} is thus a subtype of \ac{DBT}, in spite of the former usually being mistaken for the latter.

A particular subtype of \ac{DBT} is known as \enquote{blockchain.} Blockchain’s main characteristic is that transaction data is stored in data packages called \enquote{blocks} which include cryptographic references to previous blocks, thus forming a chain of blocks. All blockchains are \ac{DBT}, but not all \ac{DBT} is blockchained. However, \ac{DBT} is based on/inspired by blockchain. Therefore, blockchains and \ac{DBT} are in the same industry category.\footnote{This definition is not shared by all. Odom (personal communication, February 27, 2020), for instance, claims that blockchains and \ac{DLT} are the same, and that they are necessarily PoW-based decentralised systems: \enquote{There is no such actual thing as a \enquote{\ac{DLT}} unless it includes proof-of-work. Any such project without proof-of-work is actually under centralised control, though it may contain \enquote{distributed} elements in its design.}}

The glossary table (see below) describes these categories in a more systematic manner.
\begin{figure}[H]
  \centering
  \includegraphics[width=\textwidth]{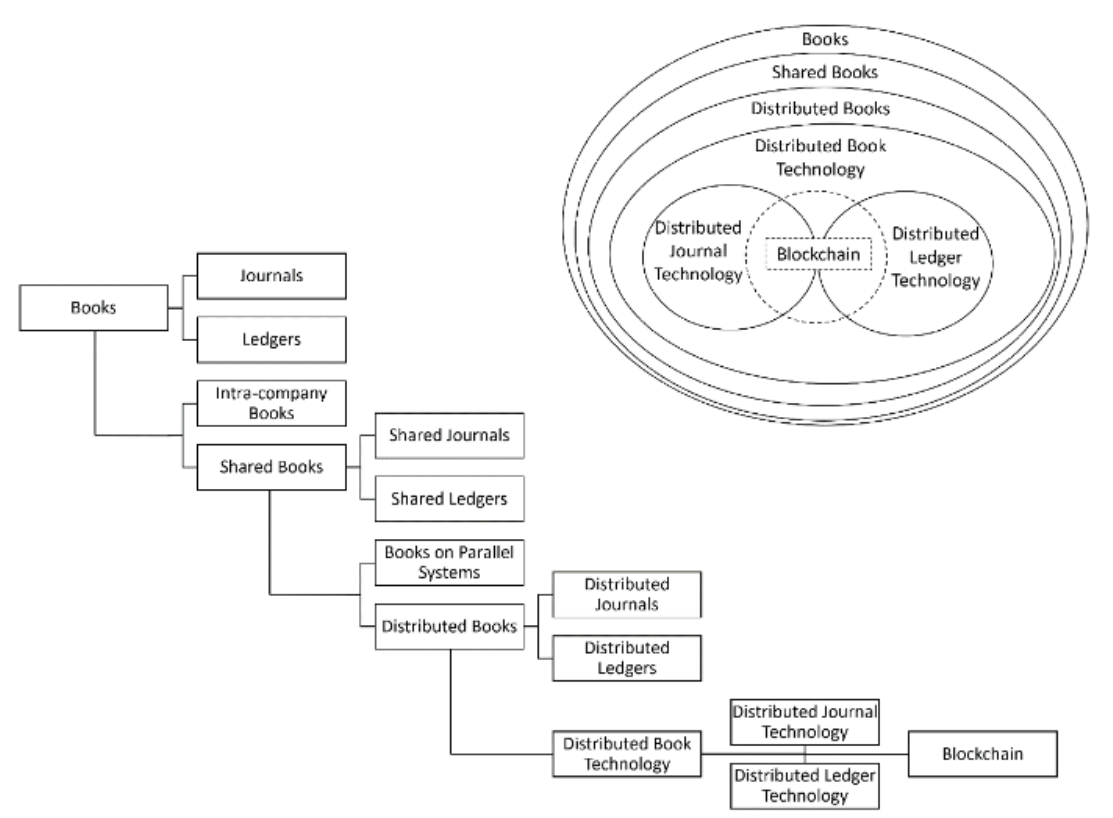}
  \caption{A representation, from an accounting perspective, of the \enquote{family tree} of categories of \ac{DLT}, the technology to which \ac{TEA} systems usually resort to (though in principle it is not required)}
  \label{figureTEA.3}
\end{figure}

\subsection{The Concept of Triple-entry Accounting}
The purpose of this section is threefold. First, to describe \ac{TEA} as a three-dimensional (3D) accounting device which comes to address certain problems in traditional, redundant, double-entry accounting (as it was originally described by Todd Boyle). Second, we define what \ac{TEA} is. Third, we enumerate the essential and nonessential features of a \ac{TEA} system.
\subsubsection{Three-dimensional Accounting}

Although there are at least two parties to a transaction,  the transaction is just one economic event. However, the parties keep historically independent and mirroring records of this event. In other words, there are two records for each one transaction. A sale, for instance, is \enquote{recorded into receivables by a seller, (...) [and] as a purchase, into accounts payable by a buyer} \citep{Boyle2000b}. Furthermore, each party records the transaction in a double-entry manner. \enquote{That's quadruple entry. Since every transaction in the developed world also causes quadruple entries when it clears [through] a bank, that's octuple entry} (\cite{Boyle2000b}; italics are ours). We can call this \enquote{redundant book-keeping\index{bookkeeping}.}

Any triple-entry system is an interparty book-keeping\index{bookkeeping} system which eliminates this redundancy: there is a shared transaction repository keeping a single record of the transaction. \cite{Boyle2003a, Boyle2003k}, see also \cite{Ibanez2023REASystems}, explains that, in this context, book-keeping\index{bookkeeping} becomes at least 3D, because there is a need to represent transactions from the viewpoint of its two parties (even if stored in a viewpoint-independent manner): the sending side and the receiving side. Attempts at representing three dimensions in two dimensions will necessarily result in redundancy, that is, recording the same transaction over and over (e.g. different, mirroring books for each party to the transaction).

\begin{figure}[H]
  \centering
  \includegraphics[width=\textwidth]{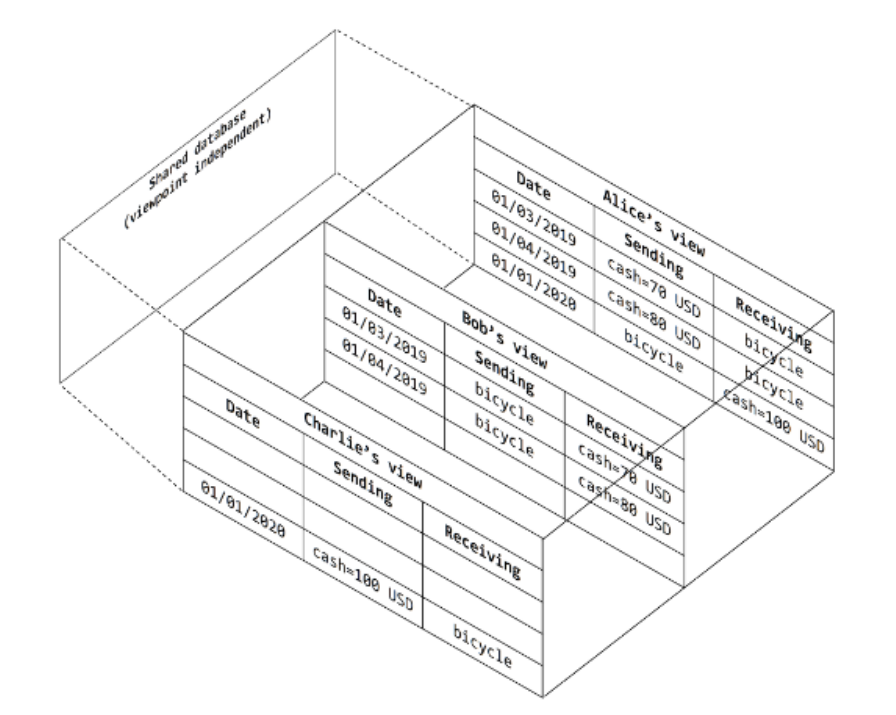}
  \caption{Three-dimensional accounting, in which N sheets equal the N parties to the system (\cite{Ibanez2023REASystems}; based on \cite{Boyle2003a,Boyle2003k}). }
  \label{figureTEA.4}
\end{figure}

Historically, 3D accounting records were not available. At present, however, computer-based shared ledgers can maintain a shared record in a 3D manner. \cite{Ibanez2023REASystems} provide a representation of 3D\footnote{More dimensions can be added, resulting in a hypercube \citep{Boyle2003a}.} accounting as explained in \autoref{figureTEA.4}.

\subsubsection{Triple-entry Accounting: A Concept}
\ac{TEB}\index{bookkeeping} is a particular proposal for a shared transaction record (and thus, for 3D accounting). It relies on signed receipts to reach an agreement on the content of the shared record. As explained by \cite{Ibanez2023REASystems}:
\begin{quotation}
    \enquote{In order to update the shared record with a new transaction record, two parties need to be involved: one initiates a transaction entry – called \enquote{request,} \enquote{offer} or \enquote{transaction draft} – and the other accepts it. This can be seen as a signature-gathering process: one party adds their signature to the transaction entry draft and the counterparty accepts by countersigning, before the entry gets processed by the STR which can be a middleware server or a distributed ledger; the STR checks the validity of the signatures and then, if everything is in order, signs off on the entry (…). This generates a hashed triple-signed receipt, such that all the parties hold the same data that cannot be manipulated or lost: a single, shared entry serving as the single source of truth.}
\end{quotation}

The two fathers of \ac{TEB}\index{bookkeeping}, Ian Grigg and Todd Boyle, conceived this sort of three-pronged consensus\index{consensus} mechanism. However, it is worth noting that only Grigg named the system \enquote{triple-entry} because of the tripartite consensus\index{consensus} involving three signed messages \citep{grigg2005triple,Ibanez2023REASystems}. \cite{Boyle2001b, Boyle2001f} called it in this manner because he had envisioned two (optional) private transaction stubs on top of the shared entry for the parties to insert nonessential data\footnote{For a discussion of the many possible meanings of the word \enquote{entry} and whether \ac{TEA} should be called \enquote{single-entry} instead of \enquote{triple-entry,} see \cite{Ibanez2023REASystems}.}. \citep{Ibanez2022Triple-entryBlockchain}

As anticipated, based on the prior distinction between book-keeping\index{bookkeeping} and accounting, we can conceive something beyond a \enquote{mere} shared record of transactions: A triple-entry system with an accounting solution. This constitutes \ac{TEA}. \ac{TEB} requires a shared journal, whereas \ac{TEA} requires either a shared ledger or a shared journal enabling individual ledgers. Given that accounting presupposes book-keeping\index{bookkeeping} (see \citealt{Ibanez2023REASystems}), but the reverse is not true, all \ac{TEA} systems are \ac{TEB} systems, but not all \ac{TEB} systems have an accounting layer.

Note that a \ac{TEA} system is not bound to any particular technology, but most current \ac{TEA} systems resort to \ac{DLT}.  Bitcoin\index{Bitcoin} is a well-known triple-entry system using \ac{DBT} \citep{Ibanez2022Triple-entryBlockchain} but, since it does not have an accounting layer, it is not \ac{TEA} and does not involve a distributed ledger. Strictly speaking, Bitcoin\index{Bitcoin} is a \ac{TEB} system with \ac{DJT} technology.

Note too that the three entries in \ac{TEA} are not three mirroring records of a transaction, but three signature entries or signed messages. This means that \ac{TEA} records may be compatible with both a single-entry and a double-entry representation of transactions.\footnote{For a more detailed discussion of the polysemy of the term \enquote{entry,} see \cite{Ibanez2023REASystems}.}

\begin{figure}[H]
  \centering
  \includegraphics[width=\textwidth]{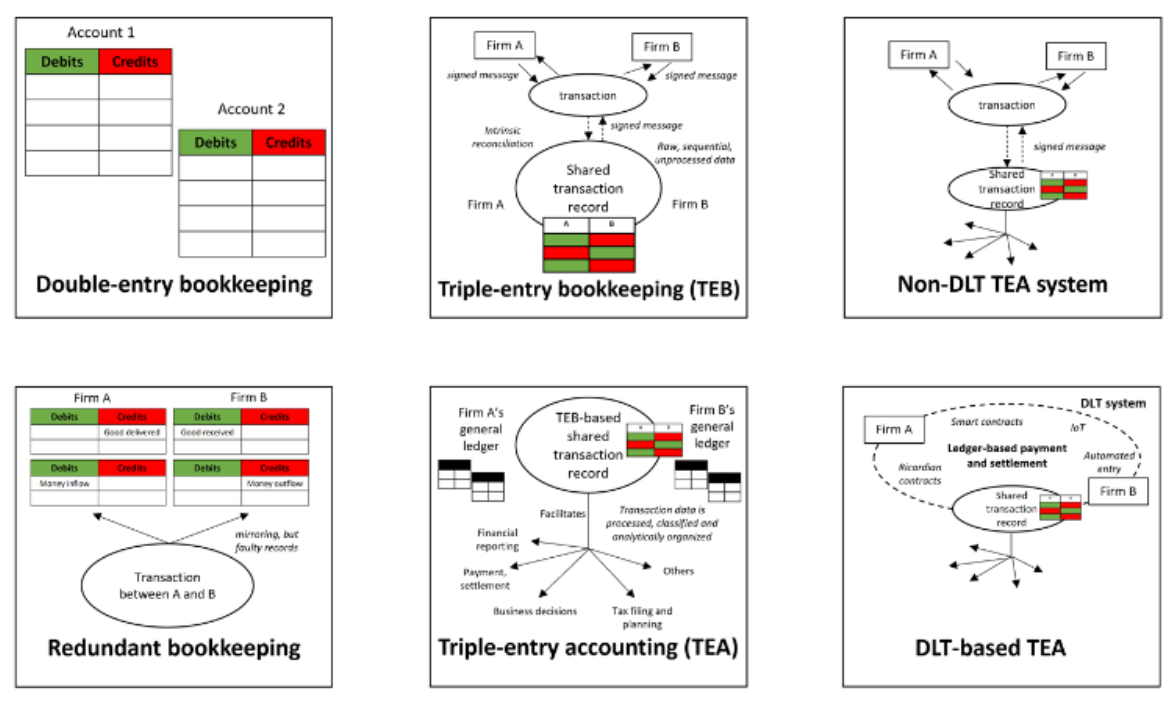}
  \caption{\ac{DLT}-based \ac{TEA} systems in comparison with other categories. Note that \ac{TEA} systems not only record, but also enable the transaction, that the record is shared between the parties (intrinsic reconciliation, unlike traditional, redundant book-keeping\index{bookkeeping}), and that transaction data is processed and organised to facilitate business life.}
  \label{figureTEA.5}
\end{figure}

\paragraph{Criteria for \ac{TEA}}
With the categories discussed in the previous subsection in mind, we can establish that a \ac{TEA} system has intrinsic requisites and two near-essential features, numbered below. Furthermore, we have identified 14 important characteristics that a \ac{TEA} system may have, despite them not being definitionally essential. 

\subparagraph{\ac{TEA} – Operational Criteria}
\begin{enumerate}[label=(\alph*)]
\item \underline{Shared transaction record (single truth)}: a \ac{TEA} system keeps a record of the transactions between two or more unrelated parties in a shared journal or ledger. This condition is informed by the \textit{What You See Is What I See (WYSIWIS)} principle\footnote{The WYSIWIS principle does not preclude the possibility of restricted viewing permissions, as long as the information that both parties have still emerged from a central, shared record.}.
\item \underline{Three-pronged consensus}\index{consensus}: either in a digital cheque model through three signed messages resulting in a triple-signed receipt (sequential offer-acceptance-validation structure) or in a digital cash model with asynchronous acceptance.
\item \underline{Ledger / accounting layer}: transactions should not just be recorded in an unprocessed manner. Raw transaction data should be fed to each party’s general ledger or it should be processed in such a way that it facilitates corporate life (decision-making, financial analysis and forecasting, invoicing, tax planning, financial reporting, etc.).
\end{enumerate}

\subparagraph{Near-essential features}
\begin{enumerate}[label=(\alph*)]
\item \underline{Immutability of transactions}: while a shared record solves the problem of redundant/duplicate records, it creates a security challenge (on top of the pre-existing possibility of tampering). This is the need to trust that the other party will not alter the dominating record in their favour. Immutability of transactions fixes this problem.

In principle, a \ac{TEA} system without this security feature is conceivable. However, Grigg’s three-way consensus\index{consensus} through digitally signed receipts ensures immutability by design. Furthermore, the low degree of trust characterising businesses since before the first double-entry systems, makes the immutability of the shared record of transactions such a practical imperative that it can be regarded as an essential requisite. This has historically been the case\footnote{\cite{grigg2005triple} called for a secure, \enquote{reliable,} \enquote{bullet proof accounting system,} to \enquote{keep a record safe} from unconsented changes, fraud and theft. \enquote{Computer science introduces concepts such as transactions, which are defined as units of work that are atomic, consistent, isolated, and durable} (\cite{grigg2005triple} italics are ours). \cite{Boyle2000e} also stated that the economics of a public transaction record depend ‘on achieving security, which is why he proposed a \enquote{true, unforgeable record}'. See also \cite{Pacio2020}.}.

Note that since no system can be regarded to be absolutely immutable, the word \enquote{immutability} should be applied within reasonable bounds to systems that are significantly harder to modify than their competitors. Furthermore, as noted in \cite{ibanez2021efficiency}, immutability should not be absolute as it is also necessary to acknowledge that business process are often tentative and unfinished.

\item \underline{Digital character}: it is hard to maintain an analogical shared transaction record unless one of the parties is willing to entrust the other with the maintenance of the record. Because \ac{TEA} is itself a concept designed for this kind of trust not to be needed, a digital system results a practical imperative.

\item \underline{Digital identity verification}: although Grigg claimed that digital cash systems do not depend on identity \citep{Saia2021TaxonomyReview}, the \ac{TEA} whitepaper proposed binding identity in a Ricardian contract.  \cite{Boyle2003j} argued that a \ac{TEA} system requires at least an alias, which means it cannot be a completely anonymous system\footnote{\cite{Boyle2003h} went beyond to state that there is also a practical requisite to comply with anti-terrorism and anti-money laundering legislation.}. In practice, most  \ac{TEA} systems will need to consider identity for practical reasons.
\end{enumerate}

\subparagraph{Architectural Add-ons}
\begin{enumerate}[label=(\alph*)]
\item \underline{Network-based settlement/payment}: in theory, a\ac{TEA} system can function as a mere archive, recording transactions common to two or more parties that conduct their dealings elsewhere. However, if the network both enables and records payment, an enhancement in efficiency may be achieved \citep{Boyle1999, Boyle2001f, Boyle2003b}

\item \underline{Smart contracts-enabling network}: network-based settlement/payment does not presuppose smart contracts, but the latter might increase the value of the former feature by making the network more versatile. Furthermore, smart contracts may play a role in automating accounting practices \citep{Ibanez2023REASystems}.

\item \underline{Ricardian contract-enabling network}: neither network-based settlement/payment nor smart contracts imply Ricardian contracts, but the latter might enhance the appeal of the former by improving the legal standing of network-based transactions.

\item \underline{Distributed Ledger (more than one node)}: the shared ledger is set up on a system of distributed computing, that is, with independent nodes communicating through message passing.
\begin{itemize}
    \item \underline{\ac{DLT}}: the distributed ledger is a decentralised, blockchained, or blockchain-inspired system for nodes lacking mutual trust.
    \begin{itemize}
        \item \underline{Blockchain (permissioned or otherwise)}: a blockchain is not necessary to enable, notify, store, and timestamp transactions. However, it might prove to be the most efficient method to do all of the above.
    \end{itemize}
\end{itemize}

\item \underline{\enquote{Stub – shared entry – stub} structure}: to enable the recording of information that is non-essential from the shared perspective, but
that could be very important for each party’s internal accounting.

\item \underline{General ledger for transactions}: a \ac{TEA} system need not involve a master record of a firm’s transactions. Nevertheless, if it did so, the system would be in a position to compete with other general ledger software accounting solutions (see \citealt{ibanez2021efficiency}).

\item \underline{General ledger for reporting}: to facilitate financial reporting, tax reporting, sales reporting, etc., is not an essential feature of a \ac{TEA} system, but its presence may improve the system’s appeal.

\item \underline{Viewing permissions}: a \ac{TEA} repository with recorded transactions may be public (e.g., gossip protocols) or private. A private \ac{TEA} could, for example, be encrypted, only allowing people with access rights to view the transactions \citep{Boyle2000c, Boyle2001b, Boyle2003b, Boyle2003e}. However, making this optional for each transaction, \enquote{according to entity preference}, would constitute a valuable feature \citep{Pacio2020}.

\end{enumerate}

\subparagraph{Other features}
\begin{enumerate}[label=(\alph*)]
\item \underline{Scalability and high throughput}: to be viable to meet business needs in the modern world \citep{barett2019poc, Pacio2020}.

\item \underline{Compliance with financial regulations}\index{regulation}: to improve the appeal of the system \citep{Pacio2020}.

\item \underline{Affordability}: so that even small businesses can use the system \citep{Auditchain2018,Pacio2020}.

\item \underline{User-friendliness}: so that even small businesses can use the system  \citep{Boyle2000a,hildebrand2018strikingly,Pacio2020}.

\end{enumerate}

\subsection{Glossary Table}
Having defined \ac{TEA} and explicated the concepts necessary to apprehend the specificity of the notion and its implementation, we can proceed to summarise some of the most fundamental terms to explain \ac{TEA}.

\begin{longtable}{p{4cm}|p{12cm}}
    \toprule
    \textbf{Book} & A record of transactions. It may be either a journal or a ledger. \\
    \midrule
    \textbf{Journal} & The book of first entry where raw (unprocessed) transaction data is recorded sequentially. \\
    \midrule
    \textbf{Ledger} & The book of second/final entry to which the transactions recorded in a journal are processed and imported (\enquote{posted}), but in an analytical order. \\
    \midrule
    \multirow{1}[2]{*}{\textbf{General Ledger}} & A master collection of books containing all of a company’s transactions (usually in a summarised form, as subsidiary ledgers – subledgers – contain the details). It is the central transaction repository. In software applications, this repository is often split in two modules \citep{Boyle2003g}: \\
    \multicolumn{1}{r|}{} & • The \ac{GLT}, where objectively verifiable transaction data is posted; and \\
    \multicolumn{1}{r|}{} & • The \ac{GLR}, where the architecture of the ledger is reorganised through subjective adjustments to serve downstream integration and reporting requirements. \\
    \midrule
    \textbf{Shared Book} & A shared journal or shared ledger. \\
    \midrule
    \textbf{Shared Journal} & A journal maintaining transaction records for two or more unrelated parties. \\
    \midrule
    \textbf{Shared Ledger} & A ledger maintaining transaction records for two or more unrelated parties. \\
    \midrule
    \multirow{2}[2]{*}{\textbf{\makecell{Shared Transaction \\Repository (STR)\footnote{Originally named ‘public transaction repository’ or PTR \citep{Boyle2000e}.} }}} & A particular concept of a shared book by \cite{Boyle2003j}, allowing any party to any inter-company transaction to: \\
    \multicolumn{1}{r|}{} & 1. post a copy of (the time, content and parties to) the transaction; \\
    \multicolumn{1}{r|}{} & 2. notify the other party; \\
    \multicolumn{1}{r|}{} & 3. enable the counterparty to post an unforgeably timestamped, non-repudiable acceptance of the transaction; and \\
    \multicolumn{1}{r|}{} & 4. provide persistent storage of the (single history of) transactions to achieve intrinsic, robust reconciliation. \\
    \multicolumn{1}{r|}{} & While originally conceived as a middleware server with an operator \citep{Boyle2001f} and in consecutive steps, it may well be set up in a decentralised system, and the steps may well collapse into one. The STR does not need to allow for payment and settlement within the network, but it probably only makes economic sense if that is the case \citep{Boyle2001f} because otherwise it will not be able to replace the banking system \citep{Boyle1999, Boyle2003b}. \\
    \midrule
    \multirow{1}[2]{*}{\textbf{Distributed Book}} & A journal (transactions are unprocessed and sequentially organised) or ledger (transactions are processed, consolidated and analytically organised) set up on a network of computers conforming to a distributed system (instead of a parallel system). \\
    \multicolumn{1}{r|}{} & For both distributed journals and distributed ledgers, the distributed database may be, in order of decentralisation\footnote{See also \cite{grigg2016}: \enquote{Todd Boyle’s concept could not work without a server, a quorum or a blockchain.}}:  \\
    \multicolumn{1}{r|}{} & 1. a traditional database in which a computer stores a master database which is then duplicated through the network (in discrete intervals), requiring trust on all the nodes, that is, in the network operator; \\
    \multicolumn{1}{r|}{} & 2. a traditional database in which specialised software tracks changes over all the databases and replicates them across the network (on an ongoing basis), requiring trust on all the nodes, that is, in the network operator; and \\
    \multicolumn{1}{r|}{} & 3. a \ac{DBT} system. \\
    \midrule
    \textbf{\acf{DBT}} & An industry term for a particular kind of distributed book--usually either blockchain or blockchain-inspired technology: a record of transactions stored in a database set up over a network of computers/nodes which cannot trust each other \citep{brown2016distributed},\footnote{One could further distinguish between fully trustless DBT systems and walled garden DBT systems with a gatekeeper \citep{grigg2017a}} and thus resort to verification/validation (and time-stamping with unique cryptographic signatures) to maintain a consensus\index{consensus} about (the single set of) shared facts. In other words, the network operator does not maintain the records by itself (it is not \enquote{central}). \\
    \midrule
    \textbf{\acf{DJT}} & A \ac{DBT} system storing raw transaction data, sequentially and in a journal. \\
    \midrule
    \multirow{2}[2]{*}{\textbf{\makecell{Distributed Ledger\\Technology (DLT)}} }& A \ac{DBT} system going beyond first entry to processing, consolidating and organising transactions, that is, a \ac{DBT} system with a ledger instead of a journal. \\
    \multicolumn{1}{r|}{} & The term \ac{DLT} is often incorrectly applied to \acp{DJT}, despite there being no shared ledger, but only a shared journal. \\
    \midrule
    \multirow{8}[2]{*}{\textbf{Blockchain}} & A particular kind of \ac{DBT} \footnote{Using the terms \enquote{blockchain} and \enquote{DLT} interchangeably is a common practice. Many choose to do this despite being aware of the distinction, in order to conform to popular usage \citep{Aste2017,tholen2019there}. We do not do this. Instead, we always stick to the specific meaning of each term (except in literal quotes).} (\ac{DJT} or \ac{DLT}), managed by a \ac{P2P} network, storing data in packages called \enquote{blocks}\footnote{In proof-of-work systems such as Bitcoin\index{Bitcoin}, a block includes a concatenation of transactions (the set is determined by the wish of the miner seeking to verify that set of transactions), the answer to a cryptographic puzzle called \enquote{nonce,} and the hash.}, each with a unique \enquote{hash} (cryptographic signature) logically dependent on the hash of the previous block to which it is connected, thus forming a linear chain of blocks. A new block can only be added after cryptographic verification by the network, which ensures that there is only one chain of blocks, and thus, a single set of shared facts. \\
    \multicolumn{1}{r|}{} & A blockchain may be \citep{carlyle2020public,tholen2019there}: \\
    \multicolumn{1}{r|}{} & Open: anyone can read the information in the blocks. \\
    \multicolumn{1}{r|}{} & Permissionless: anyone can also write on the blocks and take part in the verification process. \\
    \multicolumn{1}{r|}{} & Permissioned: permission is required for writing and verifying. \\
    \multicolumn{1}{r|}{} & Closed: only authorised parties can read. \\
    \multicolumn{1}{r|}{} & Consortium: (only) authorised parties can write and verify. \\
    \multicolumn{1}{r|}{} & Private permissioned or \enquote{enterprise}: only the network operator can write and verify. \\
    \midrule
    \textbf{Triple-entry bookkeeping (TEB)}\index{bookkeeping} & Shared book built through signed messages resulting in cite receipts. Recordkeeping is sequential and journalised. Todd Boyle’s STR, Ian Grigg’s Ricardo Payment System and Satoshi Nakamoto's\index{Satoshi Nakamoto}  Bitcoin\index{Bitcoin} are examples of TEB. \\
    \midrule
    \multirow{4}[2]{*}{\textbf{\makecell{Triple-entry\\Accounting(\ac{TEA})}}} & A TEB system with an accounting layer, i.e. at least one of the following two characteristics: \\
    \multicolumn{1}{r|}{} & The STR is not limited to just sequentially storing transactions, but also to classifying and interpreting them, facilitating decision-making, financial analysis and forecasting, tax planning and financial reporting \footnote{While \enquote{book-keeping\index{bookkeeping}} consists in recording transactions, \enquote{accounting} consists in all of the above.}.  \\
    \multicolumn{1}{r|}{} & As in \cite{Boyle2003j} original formulation, the STR serves as a subledger for each of both parties to a transaction’s general ledger: both general ledgers retrieve the transaction data from the STR. \\
    \multicolumn{1}{r|}{} & It involves a shared ledger underpinned by triple-signed receipts formed through offer, acceptance and validation. \\
    \midrule
    \textbf{Bitcoin}\index{Bitcoin} & A (currently hybrid) P2P payment system constituting both a permissionless public blockchain and an \enquote{asynchronous} TEB system \citep{Ibanez2023REASystems} without an accounting layer. \\
    \midrule
    \textbf{Resources, \newline{}Events, \newline{}Agents (REA)} & An ontology \footnote{\enquote{A number of abstractions that generalise business events}\citep{Boyle2000b}.
} for an accounting system which replaces the classical double-entry system with an information system integrated to all functional areas of an enterprise, not just limited to the accounting department. For this purpose, it proposes a single, shared record of transactions. When applied to inter-company transactions\footnote{REA was mainly conceived for intra-business organisation, but is applicable to intercompany transactions. A number of works have developed this possibility, applying it to supply chain management \citep{haugen2000rea}}, it is compatible with \ac{TEA} and may even serve to supplement \ac{TEA}, to which it is genealogically related \citep{Ibanez2023REASystems}. \\
    \midrule
    \multirow{2}[2]{*}{\textbf{\makecell{Open-edi \\Distributed \\Business \\Transaction \\Repository \\(OeDBTR)}}} & A term within the REA ontology for a single-entry system which tracks the immutable history of changes triggering changes of state in multiple business entities, relying on the independent view of the transaction as a single source of truth and the open-edi electronic data interchange standard \citep{mccarthy2019blockchain}. \\
    \multicolumn{1}{r|}{} & An OeDBTR does not require \ac{DLT}, but \ac{DLT} \enquote{exhibits all of the principles of the distributed repository in which to store business transactions} \citep{Holman2019}. \\
    \midrule
    \textbf{Momentum \newline{}accounting} & An accounting system which, aside from wealth and income (the rate of change in wealth), records the rate of change in income. Its inventor, \cite{ijiri1982triple}, named it \enquote{triple-entry book-keeping\index{bookkeeping}} but it is really accounting\footnote{At present, both Ijiri’s and Grigg’s notions of triple-entry are used, but Grigg’s definition has found wider adoption \citep{groblacher2019triple}, particularly in cryptographic circles.
} \citep{Grigg2017b, Grigg2019}. Despite other works of Ijiri that had a minor indirect impact on the \ac{TEA} concept discussed in this chapter, momentum accounting is almost unrelated to it \citep{Ibanez2023REASystems}. \\
    \midrule
    \textbf{Russian triple-entry} & Also known as \enquote{triple-book system,} it is an accounting system which proposes a continuous update of inventory and the usage of only three books: capital book, systematic accounts book, and balance book \citep{faccia2020blockchain,platonova2016fv}. It is also unrelated to \ac{TEA}. \\
    \midrule
    \textbf{Signature} & Any token that attests an agreement at a particular point in time. \\
  \label{tab:addlabel}%
\end{longtable}
\subsection{Other Double-entry Accounting Iterations}
A number of other digital means seek to enhance conventional double-entry accounting and are worth mentioning:

\subsubsection{\acf{IDEA}}
The company \cite{Pacio2020} has proposed a concept of blockchain-based double-entry accounting adding a layer of immutability to “classic” double entry accounting (CDEA). It would constitute a better fit than \ac{TEA} for \enquote{single-entry counterparty (inter-entity) transactions,} which \enquote{do not cover the internal transactions such as depreciation, payroll, etc.} (ibid: 5). \ac{IDEA} is different from other \ac{DLT}-based accounting services in that the system should be \enquote{readily upgradeable to \ac{TEA},} so as to serve as an interim step to it. This requires the ability to scale.

\subsubsection{\acf{PTA}}
\ac{PTA} is a type of accounting software aimed at making accounting less convoluted, more accessible, more easily understood and less subject to intellectual property constraint. It gets its name from the fact that it relies on plain .txt files -- rather than, for example, SQL files. In order to achieve this, accounting data is stored in a simplified and human-readable way \citep{Michael2019}.

\ac{PTA} has a number of relations to \ac{TEA}. First, some of its advocates maintain that \ac{PTA} is a necessary (or preferable) prior step to \ac{TEA} and \ac{DLT}. To \enquote{move accounting to the blockchain at this point when everyone is still using CSV files and the proprietary software of large (...) corporations (e.g.Quickbooks and Intuit) is premature. We need a common open standard first and to then build tools around that. (...) Before blockchains can/should be used for accounting in the triple-entry form, we should have a common standard for double entry book-keeping\index{bookkeeping} that is in plain text} (Buchman, personal communication, 7 January 2020).

Furthermore, just like \ac{TEA} and REA advocates, \ac{PTA} advocates point out that CDEA is viewpoint-dependent: it only includes the partial view of a transaction of the party owning the ledger. This means that in order to \enquote{model the finances of several entities simultaneously,} redundant entries are needed. The \ac{PTA} tool \enquote{Transity,} for example, allows the automatic change of viewpoints for each of the parties modelled, which constitutes a partial workaround \citep{Sieber2018TransityAccounting}.

\section{Conclusion}
\ac{TEA} is one of the most innovative concepts at the forefront of blockchain research. Its historical influence is remarkable \citep{Ibanez2023REASystems} and potential efficiency effects are copious. At present, numerous companies compete to win the race towards the development of a widely-adopted, blockchain-based \ac{TEA} system. However, the taxonomical deficit characterising this field of research impedes apprehension of what \ac{TEA} even is.

Due to insufficient usage of accounting vocabulary in blockchain research, it is often unclear what \enquote{three entries} means in this context, whether Bitcoin\index{Bitcoin} is a \ac{TEA} system, what the difference is between Bitcoin\index{Bitcoin} and a \ac{TEA} software, whether blockchains are necessarily \ac{TEA} or vice versa and why, if \ac{TEA} is a proposal for a shared ledger system, \ac{TEA} records are different from other shared ledgers. This further reveals that the terms \enquote{shared ledger,} \enquote{distributed ledger,} \enquote{decentralised ledger} and \enquote{distributed technology} are often used imprecisely, creating difficulties for the understanding of the underlying technologies. 

With the aim of filling this gap, in this chapter, we set out to achieve a number of goals. First, we provide a proposal for a terminological standardisation that allows us to give a proper account of the \ac{TEA} phenomenon. Notably, we lay out the basic characteristics of a \ac{TEA} system. We also distinguish \enquote{bare} triple-entry systems or \ac{TEB}\index{bookkeeping} systems (such as Bitcoin\index{Bitcoin}), from \ac{TEA} systems. While the \ac{TEA} systems reviewed are DLT-based, we further distinguish DLT technology from basic DJT technology (again, the Bitcoin\index{Bitcoin} example). As a prior step, we also argue against the widely used distinction of centralised, decentralised and distributed authored by Baran.

Our work constitutes an early-stage approximation, shedding light on the conceptual and linguistical complexities surrounding the emergent technology of \ac{TEA}. Although aware of the preliminary character of our terminology, we hope that it will constitute a useful contribution to the ongoing efforts in the field of \ac{TEA} and blockchain technology.

\section{Acknowledgements}
\label{acknow}
We thank the following individuals for answering our questions per email correspondence or per interview:

\begin{table}[H]
  \centering
    \begin{tabular}{l|l}
    \textbf{Individual} & \textbf{Position} \\
    \midrule
    Chris Odom & Founder of Open-Transactions \\
    David Hartley & CEO of Pacio \\
    G. Ken Holman & CTO at Crane Softwrights Ltd and former editor of ISO/IEC 15944-21 \\
    Ian Grigg & Co-founder at Solidus/Chamapesa \\
    Jason Meyers & Founder at Auditchain \\
    Robert Haugen & Developer at Mikorizal Software \\
    Todd Boyle & Founder at International Accounting Services \\
    \end{tabular}%
  \label{ack}%
\end{table}%

\newpage
\chapterbib
\end{document}